\documentclass{article}
\usepackage[english]{babel}
\usepackage{graphicx}
\usepackage{amsmath}
\usepackage{amssymb}
\usepackage{amsxtra}
\usepackage{float}

\title{Researching of magnetic cutoff for local sources of charged particles in the halo of the Galaxy}
\author{A.O. Kirichenko$^{2}$,A.V. Kravtsova$^{2}$ M.Yu. Khlopov$^{1,2,3}$, A.G. Mayorov$^{2}$\\
$^{1}$ Institute of Physics, Southern Federal University\\ Stachki 194 Rostov on Don 344090, Russia\\
$^{2}$ National Research Nuclear University MEPhI,\\ 115409 Moscow, Russia\\e-mail aokirichenko@yandex.ru\\
$^{3}$  Université de Paris, CNRS, Astroparticule et Cosmologie,\\ F-75013 Paris, France}
\begin{document}
\maketitle

\begin{abstract}
%Abstract of your contribution\ldots
Models of highly inhomogeneous baryosynthesis of the baryonic asymmetric Universe allow for the existence of macroscopic domains of antimatter, which could evolve in a globular cluster of antimatter stars in our Galaxy. We assume the symmetry of the evolution of a globular cluster of stars and antistars based on the symmetry of the properties of matter and antimatter. Such object can be a source of a fraction of antihelium nuclei in galactic cosmic rays. It makes possible to predict the expected fluxes of cosmic antinuclei with use of known properties of matter star globular clusters
 We have estimated the lower cutoff energy for the penetration of antinuclei from the antimatter globular cluster, situated in halo, into the galactic disk based on the simulation of particle motion in the large-scale structure of magnetic fields in the Galaxy. We have estimated the magnitude of the magnetic cutoff for the globular cluster M4.

\end{abstract}


% optionally

\section{Introduction}\label{s:intro}
Today, our knowledge of the chemical composition of galactic cosmic rays is being enhanced by precision experiments such as PAMELA \cite{PAMELA}, BESS \cite{BESS} and AMS-02 \cite{AMS} in near-Earth orbit. Along with common components such as protons or helium nuclei, there is no doubt about the presence of antiprotons in the cosmic ray flux, and a search for heavier anti-nuclei is also underway.

For the first time experimentally antinuclei were discovered at accelerators, which contributed to the development of theoretical models suggesting the existence of antimatter in the Universe and, in particular, in our Galaxy \cite{experiments}. According to them, antimatter is usually classified into three groups:
\begin{itemize}
\item Relic or primary.
\item Secondary.
\item From exotic sources.
\end{itemize}

There is a classical mechanism of particle production, and antiparticles, as we understand it, are born as a secondary component(for example, positrons, antiprotons, antideuterons or antihelium \cite{Moskalenko}). But nevertheless, modern models of cosmic ray generation suggest their formation and acceleration after supernova explosions in termination shocks\cite{Oliva} and when propagating in the interstellar medium, the fraction of various components in cosmic rays changes as a result of nuclear reactions with interstellar gas\cite{Moskalenko}.

 In addition, secondary particles or antiparticles are born that are initially absent from the sources, for example, positrons, antiprotons, antideuterons or antihelium.

Also, the creation of galactic antiprotons as a result of annihilation or decay of massive hypothetical dark matter particles or during the evaporation of primordial black holes is not excluded\cite{darkmatter}. In this case, the calculated fluxes can exceed the flux of the secondary component.

Antideuterons can be formed in the same mechanisms, but with a lesser probability, they have not yet been detected in cosmic rays\cite{Moskalenko}.

Antihelium was not found either: in the case of its secondary origin, the calculated ratio of the fluxes of antihelium and helium nuclei is small and does not exceed $\sim 10^{12} - 10^{14}$ \cite{Moskalenko}. Detection of antinuclei above this value would indicate the existence of primordial antimatter, preserved from the moment of the Big Bang  \cite{newBook,DolgovAM}. The creation of 4 antinucleons at once with a relatively small relative momentum in processes in exotic sources is unlikely.

Today, primary antimatter could exist in the form of antimatter domains, which are not excluded in models of inhomogeneous baryosynthesis, taking the form of clusters of anti-stars or antigalaxies \cite{Khlopov}.
\section{Globular cluster of antistars}

Based on the symmetry of matter and antimatter \cite{experiments}, it is possible to indicate the expected parameters of a globular cluster of antistars. That is, a globular cluster will have the same set of properties as an ordinary cluster of ordinary stars. This approach assumes similar initial conditions and similar evolution of antimatter and matter. One should note that the mechanisms of nonhomogeneous baryosynthesis may lead to difference in the conditions within antimatter domain and in ordinary baryonic matter. In particular, the approach \cite{Dolgov3,Blinnikov} predicts much higher antibaryon density in antimatter domains, than in surrounding baryonic matter, what leads to prediction of ultra dense antibaryonic objects in the Galaxy and their specific effects \cite{Blinnikov}. Here we follow the approach of \cite{KRS2}, which assumes similar conditions in antimatter domains as in the surrounding matter, and elaborate the prediction of antihelium flux from antistars accessible to the AMS02 experiment \cite{AMS}, which follows from this similarity.

A globular cluster of stars is a group of stars that gather in the shape of a sphere and orbit around the core of the Galaxy.
The stars turn out to be gravitationally bound, which, in fact, is the reason for such a shape of these clusters. Globular clusters are localized far from most other objects in the Galaxy - in the halo. They are much denser than open clusters, and they are also older and contain more stars.
In the Milky Way, the number of globular clusters about 150.

The birth regions of such clusters are the dense interstellar medium. However, no star formation is currently observed in globular clusters. All dust and gas have long been "blown out" from the clusters.
This confirms the opinion that globular clusters are the oldest objects in the Galaxy\cite{cluster1}.
Orbiting the outskirts of the galaxy, globular clusters take several hundred million years to complete one orbit.
At the center of the cluster, the highest density is achieved - on the order of 100-1000 stars/$pc^3$. For comparison, the density of stars near the Sun is 0.14 stars/$pc^3$.
Globular clusters have a low metallicity due to the fact that they are composed of first generation stars. Which once again confirms the opinion that globular clusters are old clusters \cite{cluster2}

\section{Cosmic  antihelium propagation in interstellar medium}
After generation and acceleration in the source, cosmic ray particles enter the interstellar medium, where they change their original trajectory, "entangled" in the magnetic fields of the Galaxy and can leave it.

The propagation of cosmic rays in the modern view is of a diffusion nature. The GCR confinement time before leaving the Galaxy is inversely proportional to the diffusion coefficient, i.e. decreases with increasing energy. For particles with energies of 1–2 GeV, it is $\sim 4 \cdot 10^7 $ years.
During this time, they manage to fill the halo of the Galaxy and, although the substance in the Galaxy is generally very rarefied, they also pass through a thickness of matter of about 10 $g/cm^2$. For high-energy particles, the distance traveled sharply decreases and, for example, at an energy of 10 TeV is 0.1-0.4 $g/cm^2$, and the lifetime is $\sim 10^6$ years \cite{Moskalenko, Moskalenko2}.

At present, attempts are being made to calculate the fluxes of galactic CRs. Solving this problem requires knowledge of the structure and size of the Galaxy, the location and power of the sources, the location of the Solar System, and the properties of the interstellar medium. CR propagation in the Galaxy is seriously determined by the structure of magnetic fields. The regular field lines lie in the galactic plane and approximately run along the spiral arms. The average amplitude of the field strength is (2-3)$\cdot 10^{-6}$ G. The magnetic field also exists in the halo, but its structure is not exactly known.
It should be noted that currently there is a numerical implementation of the leaky box model in the form of a set of GALPROP programs.

GALPROP is a numerical solver of the diffusion equation taking into account a detailed description of the distributions of the interstellar gas and the galactic magnetic field\cite{Moskalenko}.

The approach used in this work differs from the work of the GALPROP software package, and instead of solving the transport equation, individual particles are traced in interstellar space and we also take into account the parameters of the interstellar medium and the structure of magnetic field.

\section{Boris - Bunemann tracing method}
Now there are various software packages that perform tracing of particles in electromagnetic fields \cite{1}.
In 1970 Boris \cite{2} proposed a convenient way to solve the equations of motion of particles in electromagnetic fields, this method is now widely known as the Boris method. De facto, it is the standard for modeling particle motion in plasma.

The method solves the classical equation of motion in an electromagnetic field specified by vectors ${\vec{E}}$ and  ${\vec{B}}$ (vectors of electric and magnetic fields, respectively). Further, the electric field is eliminated by redefining the variables and the equation is reduced to describing only the rotational motion in the magnetic field. Then Bunemann introduces some additions to Boris's algorithm that increase the accuracy of the method. Today, there is an implementation of the method with the inclusion of relativistic corrections \cite{3}.
\subsection{Using of method }
To make the method convenient to use, a software package was created that allows you to transfer all the necessary parameters to the function for calculating the trajectory of the environment. Below is a list of them.

\begin{itemize}
\item Particle initial coordinates and directional distribution.
\item  Particle type and energy.
\item Configuration of magnetic and electric fields.
\item Characteristics of temporal and spatial steps for numerical solution.
\item  Conditions for saving trajectories, interrupting the tracing algorithm.
\end{itemize}
It is also important to note that for the development of the software package, it is possible to determine the interstellar medium for calculating the absorption of cosmic rays, and a program for calculating such interactions.
\subsection{Helium antinucleus tracing}
The following initial conditions were chosen for tracing helium antinuclei:
\begin{itemize}
\item The initial coordinates correspond to the globular cluster M4 with the position (-5.9, -0.3, 0.6) kpc in the galactic coordinate system \cite{1}.
\item The angular distribution of particles at the point of birth is isotropic.
\item The energy of particles varied from 10 GeV to 10 TeV.
\item The structure of regular component of magnetic field of the Galaxy is taken from publication \cite{2}.
\end{itemize}

\section{Results}
Figure 1 shows examples of trajectories of helium antinuclei launched towards the plane of the galactic disk with different energies.
\begin{figure}[htp]
\centering
\includegraphics[scale=0.44]{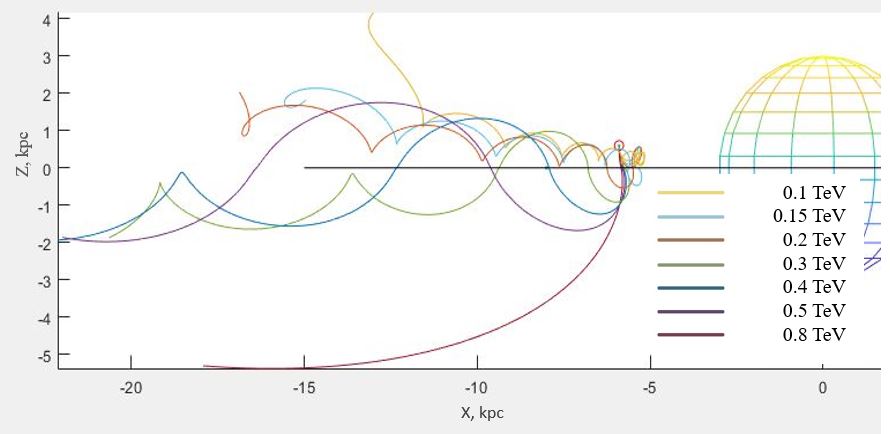}\\
\caption {Helium trajectories in regular galactic magnetic field. The plane of the galactic disc is shown in black.}
\label{monoline}
\end{figure}
Particles with an energy of 100 GeV did not penetrate into the plane of the disk, deflected and flew away into intergalactic space. Particles of higher energy penetrated into the galactic disk and at an energy close to 1 TeV they had the opportunity to leave it.

In Fig. 2, the line connecting the points with errors shows the fraction of antihelium that fell into the galactic disk 300 pc thick, depending on the particle energy.
\begin{figure}
\centering
\includegraphics[scale=0.45]{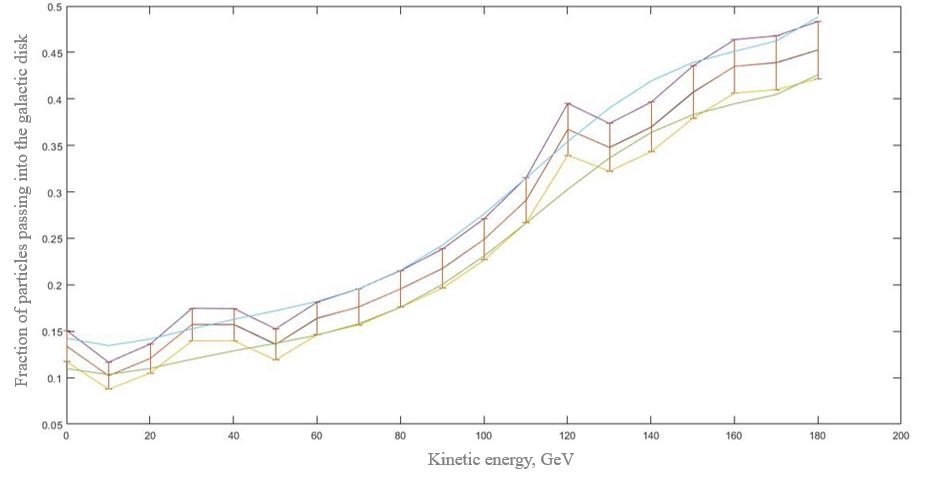}\\
\caption{The dependence of the fraction of particles penetrated into the disk on the particle energy}
\label{monoline}
\end{figure}

The obtained dependence was smoothed taking into account the error and the energies were determined at which the smoothed curves intersect the 0.25 level, i.e. width at half-height of the graph (with increasing energy, the graph tends to a value of$ \sim 0.5$, which corresponds to the geometric factor of the disk plane from the point with the coordinates of the M4 cluster). The obtained energy -- magnetic cutoff energy is 100 $\pm 10$ GeV. This means that the flux of helium antinuclei from the hypothetically globular cluster of antistars M4 will be largely suppressed at energies less than $ \sim 100$ GeV, but will not be completely suppressed.
\section{Conclusion}
Helium antinuclei were traced with \cite{Golubkov} from the M4 cluster, hypothetically consisting of antistars, to the plane of the galactic disk. The characteristic energy of magnetic cutoff is determined, below which it is difficult for particles to penetrate into the disk. Predictions of antihelium flux would strongly depend on the interference of the initial spectrum, which is expected to be falling down with energy and magnetic cutoff, which redices the lower energy part of the spectrum in galactic disk. The result will be used to interpret the experimental data on antinuclear fluxes obtained by the PAMELA and AMS-02 in near-earth orbit.

The preliminary indications to possible detection of antihelium events in AMS02 experiment, which cannot be explained as secondaries from astrophysical sources \cite{poulin}, if confirmed, would become serious evidence for existence of forms of primordial antimatter in our Galaxy. It will favor Beyond the Standard Model (BSM) physics, which can support creation and survival of antimatter domains in baryon asymmetrical Universe, and provide a sensitive probe for parameters of the corresponding models \cite{PPNP}. Whatever is the actual form of antimatter objects in our Galaxy, propagation of antinuclei from these sources would inevitably involve their diffusion in galactic magnetic fields, studied in the present paper.
\section*{Acknowledgements}
The work by AK and AM has been supported by the grant of the Russian Science Foundation (Project No-18-12-00213-P).

\end{document}